\documentclass[a4paper]{article}
\usepackage{a4wide}
\usepackage{graphicx} \usepackage{amsmath,amssymb,amsthm}
\usepackage{hyperref}
\usepackage[ruled,vlined,linesnumbered]{algorithm2e}
\theoremstyle{remark}
\newtheorem{definition}{Definition}
\newtheorem{fact}{Fact}
\newtheorem{lemma}{Lemma}
\newtheorem{corollary}{Corollary}
\newtheorem{theorem}{Theorem}

\DeclareMathOperator{\LCE}{\textsf{lce}}
\DeclareMathOperator{\LCS}{\textsf{lcs}}
\DeclareMathOperator{\nxt}{\textsf{nxt}}

   \SetKwFunction{Fill}{\textsf{fill}}
\usepackage{longtable}
\usepackage{needspace}
\usepackage[capitalise]{cleveref}
\crefname{fact}{Fact}{Facts}
\Crefname{fact}{Fact}{Facts}

\def\problembox#1{\vspace{2mm}\noindent\fbox{\begin{minipage}{.985\textwidth}#1
    \end{minipage}}\vspace{2mm}}

\title{Output-Sensitive Enumeration of Maximal Closed Substrings in Linear Space}
\author{Johannes Fischer and Dominik Köppl}
\date{}

\begin{document}
\maketitle

\begin{abstract}
We revisit the output-sensitive enumeration of maximal closed substrings (MCSs). 
Given a string of length $n$ over an integer alphabet, 
let $k$ denote its number of MCS occurrences. 
We combine a right-to-left sweep on the string with range-successor queries to obtain all MCSs with two deterministic word-RAM bounds: 
for any fixed $\varepsilon>0$, 
we can report all MCSs in $O(n\sqrt{\log n}+k\log^\varepsilon n)$ time using $O(n)$ words of space,
or in $O(n\sqrt{\log n}+k\log\log n)$ time using $O(n\log\log n)$ words. 
Both bounds include preprocessing and report occurrences by their endpoints. 
In particular, these algorithms improve the known $O(n\log n)$ worst-case time bound for sufficiently small output.
\end{abstract}

\section{Introduction}
A string is \emph{closed} if it is a single character or has a nonempty border occurring only as its prefix and suffix. 
A \emph{maximal closed substring} (MCS) is a closed occurrence that cannot be extended by one character to either side while preserving closedness. 
MCSs capture runs and certain maximal gapped repeats. 
Badkobeh et al.~\cite{badkobeh2022mcs} introduced MCSs and gave an output-sensitive $O(n\log n+k)$-time algorithm for constant-size alphabets; 
their subsequent journal version~\cite{badkobeh2026mcs} achieves $O(n\log n)$ worst-case time.
Kosolobov~\cite{kosolobov2024closed} characterized MCSs equivalently as closed repeats and established that their worst-case number is $\Theta(n\log n)$, 
already for binary strings. 
Jain and Mhaskar~\cite{jain2026closed} developed compact representations of all closed substrings and additional $O(n\log n)$-time MCS algorithms.
Two recent SPIRE~2026 papers address different computational settings:
Shibata et al.~\cite{shibata2026online} obtain an $O(n\log n)$-time, $O(n)$-space online algorithm using a link-cut suffix tree, 
whereas Umezaki et al.~\cite{umezaki2026rle} enumerate a compact family representation of MCS occurrences directly from run-length encoded input. 
Our focus is instead on output-sensitive enumeration from an uncompressed, static text.

A reason for studying output-sensitive enumeration is that the number $k$ of MCS occurrences can be much smaller than $n$. 
For example, let $T=\mathtt{a}^m\mathtt{b}^m$ with $m\ge1$ and $n=2m$.
Its only MCSs are the two full runs $T[1..m]=\mathtt{a}^m$ and $T[m+1..2m]=\mathtt{b}^m$: 
shorter substrings within a run can be extended while remaining closed, 
whereas substrings crossing the run boundary have
no nonempty border. 
We conclude that $k=2$, independent of $n$.
This contrasts with the worst-case $\Theta(n\log n)$ output size and illustrates the benefit of reducing the input-dependent part of the running time: 
choosing $0 < \varepsilon \le 1/2$, our linear-space bound specializes to $O(n\sqrt{\log n})$ total time on this family, 
compared with the $O(n\log n)$ worst-case guarantees of the previous algorithms.

We revisit output-sensitive MCS enumeration with a right-to-left sweep that maintains
right-maximal closed-substring candidates in order of their lengths.
The algorithm groups successive candidates witnessing the same MCS into a compact
representation, which we call a \emph{manacle}, and processes creations and deletions (in the following called \emph{births} and \emph{deaths}) of manacles rather than all candidates.
With $t_{\mathrm{nxt}}$ denoting the time for a succeeding-occurrence query
$\nxt_i(h)$ with $h\ge2$, the enumeration takes
$O(n+k t_{\mathrm{nxt}})$ time after preprocessing. 
For an integer alphabet of polynomial size,
we obtain the following two end-to-end deterministic bounds, reporting
each MCS by its endpoints:
\[
\begin{array}{ll}
O(n\sqrt{\log n}+k\log^\varepsilon n)\text{ time}
 & \text{and }O(n)\text{ words of space},\\
O(n\sqrt{\log n}+k\log\log n)\text{ time}
 & \text{and }O(n\log\log n)\text{ words of space},
\end{array}
\]
for any fixed $0 < \varepsilon < 1$. 
Compared with the earlier $O(n\log n+k)$ output-sensitive bound~\cite{badkobeh2022mcs},
our input-dependent term is smaller, at the price of a polylogarithmic factor per output.
For $0<\varepsilon<1$, the first bound is $o(n\log n)$ when
$k=o(n\log^{1-\varepsilon}n)$, and the second is $o(n\log n)$ when
$k=o(n\log n/\log\log n)$.

Interestingly, the $O(n\sqrt{\log n})$-term in the time bound is similar to best linear-space solution for computing the \emph{rightmost Lempel-Ziv} parsing of a text of length $n$ and alphabet size $\le n$ \cite{belazzougui2016range}. This is no big surprise, as one can easily compute the rightmost LZ-parsing from all MCSs: if $T[i..j]=\alpha\beta\alpha$ is an MCS with maximal border $\alpha$, there is no occurrence of $\alpha$ in between, so the first occurrence $T[i..i+|\alpha|)$ of $\alpha$ is the rightmost occurrence to the left of the second occurrence $T(j-|\alpha|..j]$, allowing us to point to $j$ if there is an LZ-phrase starting at $j-|\alpha|+1$. So computing all MCSs of a string can be seen as a generalization of \emph{rightmost} Lempel-Ziv, in a similar way as the array of all \emph{longest previous factors} generalizes \emph{regular} Lempel-Ziv.

As a further side note, the distinction in computational models matters:
Kosolobov~\cite{kosolobov2024closed} proves an $\Omega(n\log\sigma)$ lower bound for closed repeats over a \emph{general ordered alphabet},
even when the output has size $O(n)$. 
It does not apply to our integer-alphabet word-RAM model.

\begin{figure}[t]
  \centering
  \includegraphics[width=0.8\textwidth]{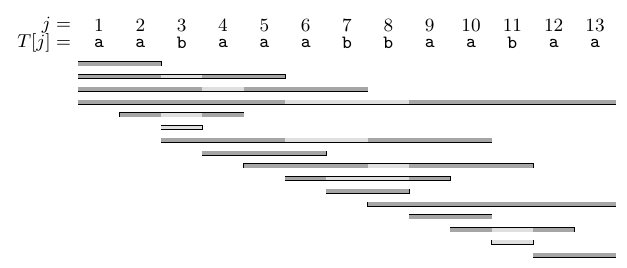}
  \caption{MCSs of $T=\mathtt{aabaaabbaabaa}$ (sentinels omitted). The nonempty longest border of each MCS is shown in dark gray.
    The substring $T[2..7] = \mathtt{abaaab}$ is closed, right-maximal, but not left-maximal since it can be extended to the left by $\mathtt{a}$ to $T[1..7] = \mathtt{aabaaab}$, which is also closed with border $\mathtt{aab}$.
  }
  \label{fig:mcs}
\end{figure}

Throughout the paper, $T=\mathtt{aabaaabbaabaa}$ serves as our running example, and
\Cref{fig:mcs} illustrates its MCS occurrences.

\section{Preliminaries}

Let $T[1..n]$ be a string over an integer alphabet $\Sigma$ of size $\sigma\le n^{O(1)}$. For
$1\le i,j\le n$ we write $T[i..j]$ for $T[i]T[i+1]\cdots T[j]$, which is the empty string
$\varepsilon$ if $i>j$. To avoid case distinctions at the two ends of the text we fix two
\emph{sentinels}
$\$_1\neq\$_2$ with $\$_1,\$_2\notin\Sigma$ such that
each sentinel occurs exactly once in $T[0..n+1]$, namely at 
$T[0]=\$_1$
and $T[n+1]=\$_2$.
A string $B$ is a \emph{border} of a string $S$ if $B$ is both a proper prefix and a proper suffix of
$S$. A string $S$ is \textbf{\emph{closed}} if $|S|\le1$, or if $S$ has a nonempty border that has exactly two
occurrences in $S$. It is easy to see that we only have to consider the longest borders for deciding on the closedness of strings:

\begin{lemma}\label{lem:longestborder}
If a nonempty border $B$ of $S$ has exactly two occurrences in $S$, then $B$ is the longest border of
$S$. In particular, the border witnessing closedness is unique.
\end{lemma}

\begin{proof}
Let $X$ be a border of $S$ with $|X|>|B|$. As $B$ and $X$ are both prefixes of $S$, $B$ is a prefix of $X$. As both are suffixes of $S$, $B$ is a suffix of $X$. 
So $B$ occurs in $S$ at position $|X|-|B|+1$, which is greater than $1$ because $|X|>|B|$ and smaller than $|S|-|B|+1$ because $|X|<|S|$. This is a third occurrence of $B$, a contradiction.
\end{proof}

Our computational task will be to detect closed substrings of a given string.
In order to constrain the output meaningfully, we say that
an occurrence $T[i..e]$ with $1\le i\le e\le n$ is a \textbf{\emph{maximal closed substring}} (MCS) of $T$ if
$T[i..e]$ is closed while neither $T[i-1..e]$ nor $T[i..e+1]$ is closed. We write $k$ for the number
of MCSs of $T$ throughout this paper.

We report an MCS occurrence by its pair of endpoints $(i,e)$.
By doing so, outputting an MCS alone takes $O(1)$ time regardless of the length of the MCS.
We are now ready to state the problem we solve in this paper.

\problembox{\textsc{MCS-Computation}\\
  \textbf{Input:} text $T[1..n]$.\\
  \textbf{Output:} all $k$ MCSs of $T$, each MCS $T[i..e]$ as a pair of endpoints $(i,e)$.
}

The substrings of length one can be detected trivially by the following fact:

\begin{fact}\label{fact:singletons}
  A character $T[i]$ is an MCS if and only if $T[i] \not\in \{ T[i-1], T[i+1] \}$.
\end{fact}

To solve \textsc{MCS-Computation} in general, we introduce the following notation.
For $1\le i,j\le n+1$, define the longest common right-extension of two text positions
\[
  \LCE(i,j)=\max\{h\ge0 : T[i..i+h-1]=T[j..j+h-1],\; i+h-1\le n,\; j+h-1\le n\},
\]
and for $0\le i,j\le n$, define the longest common left-extension of two text positions
\[
  \LCS(i,j)=\max\{h\ge0 : T[i-h+1..i]=T[j-h+1..j],\; i-h+1\ge1,\; j-h+1\ge1\}.
\]
For $1\le j\le n$ and $1\le h\le n-j+1$, we denote the \emph{succeeding occurrence} of $T[j..j+h-1]$ by $\nxt_j(h)$, formally defined as
\[
  \nxt_j(h)=\min\{q : j<q\le n-h+1,\; \LCE(q,j) \ge h\}.
\]
We define $\nxt_j(h)=\infty$ if no such $q$ exists,
and let $\lambda_j := \max\{h \ge 1 : \nxt_j(h) < \infty \}$ be the length of the longest extension starting at $j$ that has a later occurrence in the text,
where $\lambda_j := 0$ if $\nxt_j(1)=\infty$, i.e., if $T[j]$ is the last occurrence of character $T[j]$ in $T$. An occurrence of a prefix of $T[j..n]$ is also an occurrence of every shorter prefix, so $\nxt_j$ is
non-decreasing on $[1,\lambda_j]$ and $\nxt_j(h)=\infty$ for $h>\lambda_j$.

\section{Structure of Maximal Closed Substrings}
\label{sec:structure}
We first show a basic but important connection between the $\nxt$-function and MCSs:
\begin{lemma}
\label{lem:shape}
Let $1\le i<e\le n$. Then $T[i..e]$ is closed if and only if there is an $\ell$ with
$1\le\ell\le e-i$ and $\nxt_i(\ell)=e-\ell+1$. In that case $\ell$ is unique and equals the length of
the longest border of $T[i..e]$.
\end{lemma}

\begin{proof}
($\Leftarrow$) Put $p=e-\ell+1=\nxt_i(\ell)$ and $B=T[i..i+\ell-1]=T[p..e]$. Then $B$ is a border of $T[i..e]$, proper because $\ell\le e-i$.
An occurrence of $B$ inside $T[i..e]$ starting at $q$ with $i<q<p$ contradicts $p=\nxt_i(\ell)$, and an occurrence starting at $q>p$ ends beyond $e$.
So $B$ occurs exactly twice in $T[i..e]$, and $T[i..e]$ is closed.

($\Rightarrow$) Let $B$ be a border of $T[i..e]$ with exactly two occurrences and $\ell=|B|\ge1$.
Its occurrences are at $i$ and at $p=e-\ell+1$ and there is none in between, so $\nxt_i(\ell)=p$. $\ell\le e-i$ holds since $B$ is proper.

The uniqueness of $\ell$ follows easily from \cref{lem:longestborder}.
\end{proof}

The next lemma shows how closed substrings behave when extended by one character to the right or left in $T$:

\begin{lemma}\label{lem:ext}
Let $T[i..e]$ with $i<e$ be closed, let $B$ be its longest border, let $\ell=|B|$, and let
$p=e-\ell+1=\nxt_i(\ell)$. Then
\begin{itemize}
\item[(i)] $T[i..e+1]$ is closed $\iff$ $T[i+\ell]=T[e+1]$ $\iff$ $\LCE(i,p)>\ell$
\item[(ii)] $T[i-1..e]$ is closed $\iff$ $T[i-1]=T[p-1]$.
\end{itemize}
\end{lemma}

\begin{proof}
(i) If $T[i+\ell]=T[e+1]$, then $X=T[i..i+\ell]=T[p..e+1]$ is a border of $T[i..e+1]$. Any occurrence of $X$ starting at $q$ with $i<q<p$ would contain an occurrence of $B$ at $q$, contradicting $p=\nxt_i(\ell)$, and an occurrence starting at $q>p$ ends beyond $e+1$. Hence $T[i..e+1]$ is closed.

Conversely, let $T[i..e+1]$ be closed with longest border $Y$, and let $\ell''=|Y|$. If $\ell''\le\ell$, then $Y$ is a prefix of $B$ and occurs at $i$, at $p$ and as a suffix at $e-\ell''+2$.
But these are three distinct occurrences, because $p+\ell''-1\le p+\ell-1=e<e+1$. If $\ell''\ge\ell+2$, then $T[i..i+\ell''-2]=T[e-\ell''+2..e]$ is a border of $T[i..e]$ of length $\ell''-1>\ell$, contradicting the maximality of $\ell$.
We thus have $\ell''=\ell+1$, which means $T[i+\ell]=T[e+1]$.
The second equivalence follows from $\LCE(i,p)\ge\ell$ and $T[p+\ell]=T[e+1]$.

(ii) If $T[i-1]=T[p-1]$, then $T[i-1..i+\ell-1]=T[p-1..e]$ is a border of $T[i-1..e]$, and an occurrence of it at $q$ with $i-1<q<p-1$ would give an occurrence of $B$ at $q+1\in(i,p)$, which is impossible.
Conversely, let $T[i-1..e]$ be closed with longest border of length $\ell''$. If $\ell''\le\ell$, that border is a suffix of $B$, so it also occurs at $i+\ell-\ell''$, which lies strictly inside $T[i-1..e]$, giving a third occurrence. If $\ell''\ge\ell+2$, deleting the first character of both occurrences of that border yields a border of $T[i..e]$ of length $\ell''-1>\ell$, a contradiction.
So $\ell''=\ell+1$, forcing $T[i-1]=T[p-1]$.
\end{proof}

This allows us to characterize all MCSs in the following, algorithmically approachable way:

\begin{corollary}
\label{cor:char}
Let $1\le i<e\le n$. Then $T[i..e]$ is an MCS if and only if there is an $\ell\ge1$ such that, with
$p=\nxt_i(\ell)$, the following conditions hold:
\begin{itemize}
	\item   $p=e-\ell+1$,
	\item   $\LCE(i,p)=\ell$ (accounting for right-maximality), and
	\item   $T[i-1]\neq T[p-1]$ (accounting for left-maximality).
\end{itemize}
\end{corollary}

Our algorithm first finds right-maximal closed substrings and then tests whether they admit a closed left extension. We call these right-maximal substrings \emph{candidates}, formally defined as:

\begin{definition}
  For $j \in [1..n]$ the set of \emph{candidates} at $j$ is
\[
  \mathcal C_j=\bigl\{(p,\ell) : 1\le\ell\le \lambda_j,\; p=\nxt_j(\ell)<\infty,\; \LCE(j,p)=\ell\bigr\}.
\]
\end{definition}

\begin{figure}[t]
  \centering
  \includegraphics[width=\textwidth]{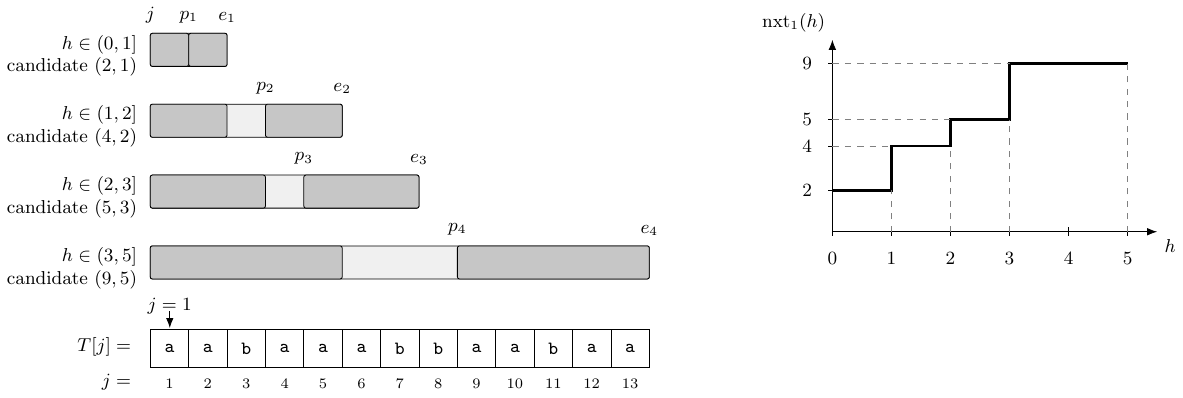}
  \caption{Candidate set $\mathcal C_1=\{(2,1),(4,2),(5,3),(9,5)\}$ at text position $j=1$ of our running example.
    \emph{Left}: Visualization of $\mathcal C_1$ in text range with MCSs starting at $j=1$ drawn as horizontal bars. The longest border of each MCS is filled in dark gray inside its bar.
    \emph{Right}: Visualization of the depth ranges in \cref{lem:staircase} as a staircase.
    The number of steps is the cardinality of $\mathcal C_1$.
    Each step spans the depth range of its candidate, while its height gives the text position of the second starting position of its longest border (the right dark gray bar).
  }
  \label{fig:staircase}
\end{figure}

Since a candidate $(p,\ell)$ witnesses the closed substring $T[j..p+\ell-1]$, 
the \emph{labels} $p_i$ and \emph{lengths} $\ell_i$ of the tuples in $\mathcal C_j$ have some structure due to the definition of closed substrings.
In detail, the next lemma shows that for increasing right endpoints $p$ of candidates at $j$, the lengths $\ell$ of borders must also be increasing:

\begin{lemma}
\label{lem:staircase}
Assume $\lambda_j>0$, and let $\mathcal C_j=\{(p_1,\ell_1),\dots,(p_m,\ell_m)\} \ne \emptyset$ with $p_1<\dots<p_m$. 
Then
$\ell_0 := 0 < \ell_1<\dots<\ell_m=\lambda_j$, and we have $\nxt_j(h)=p_t$ for all
$h\in(\ell_{t-1},\ell_t]$. We call $(\ell_{t-1},\ell_t]$ the \emph{depth range} and $\ell_t$ the
\emph{length} of the candidate $(p_t,\ell_t)$.
\end{lemma}

\begin{proof}
Let $p$ be a value attained by $\nxt_j$ on $[1,\lambda_j]$. By monotonicity, $I=\{h:\nxt_j(h)=p\}$ is an interval, say $I=(h^-,h^+]$. For every $h\in I$ the position $p$ is an occurrence of $T[j..j+h-1]$, so $\LCE(j,p)\ge h^+$. In addition, $\nxt_j(h^++1)\neq p$, either because it is larger or because it is $\infty$.
In both cases $p$ is not an occurrence of $T[j..j+h^+]$, so $\LCE(j,p)=h^+$. So the pairs $(p,h^+)$ over all attained values $p$ are exactly the candidates, and their intervals partition $[1,\lambda_j]$.
\end{proof}

See \cref{fig:staircase} for an example of the resulting structure of candidates.
By \cref{cor:char} and \cref{lem:staircase}, the candidates at $j$ correspond exactly
to the closed substrings starting at $j$ that cannot be extended to the right, and such a candidate
is an MCS if it cannot be extended to the left, either.

We now show that half the length of the suffix $T[j..n]$ bounds the number of candidates at $j$.
\begin{lemma}
\label{lem:width}
For every $1\le j\le n$ we have $|\mathcal C_j|\le\lfloor(n-j+1)/2\rfloor$.
\end{lemma}

\begin{proof}
If $\mathcal C_j=\emptyset$, the claim is immediate.
Otherwise, let $\mathcal C_j=\{(p_1,\ell_1),\dots,(p_m,\ell_m)\}$ as in
\cref{lem:staircase}. The labels satisfy $j<p_1<\dots<p_m$ and are integers, so $p_t\ge j+t$.
The lengths satisfy $1\le\ell_1<\dots<\ell_m$, so $\ell_t\ge t$.
Since $p_t+\ell_t-1\le n$, we have $j+2m-1\le p_m+\ell_m-1\le n$, and therefore $m\le(n-j+1)/2$.
\end{proof}

To see what happens to the candidate set when moving from $j$ to $j-1$, we show:

\begin{lemma}
\label{lem:shift}
Let $2\le j\le n$. Then $\nxt_{j-1}(1)=\min\{q>j-1:T[q]=T[j-1]\}$, and for $h\ge2$, with $p=\nxt_j(h-1)$,
\[
  \nxt_{j-1}(h)\;\begin{cases} =p-1, & \text{if } p<\infty \text{~and~} T[p-1]=T[j-1],\\
   >p-1, & \text{if } p<\infty \text{~and~} T[p-1]\neq T[j-1],\\
   =\infty, & \text{if } p=\infty.\end{cases}
\]
In particular $\lambda_{j-1}\le \lambda_j+1$.
\end{lemma}

\begin{proof}
Let $c=T[j-1]$ and $X=T[j..j+h-2]$. 
A position $q\ge j$ is an occurrence of $cX$ if and only if
$T[q]=c$ and $q+1$ is an occurrence of $X$. 
By minimality of $p$ there is no occurrence of $X$ in
$(j,p)$,
which means that no occurrence of $cX$ starts in $[j,p-1)$. 
Finally, if $T[p-1]=c$, then $p-1$ is an occurrence of $cX$ and therefore the leftmost one. If $X$ does not occur after $j$, then $cX$ does not occur after $j-1$.
\end{proof}

\begin{lemma}
\label{lemma:manacle}
Let $(p,\ell)\in\mathcal C_j$ with $j\ge2$. If $T[j-1]=T[p-1]$, then
$(p-1,\ell+1)\in\mathcal C_{j-1}$. If $T[j-1]\neq T[p-1]$, then $T[j..p+\ell-1]$ is an MCS.
\end{lemma}

\begin{proof}
In the first case $\nxt_{j-1}(\ell+1)=p-1$ by \cref{lem:shift}, and
$\LCE(j-1,p-1)=1+\LCE(j,p)=\ell+1$. The second case is \cref{cor:char}.
\end{proof}

\begin{figure}[t]
  \centering
  \includegraphics[width=\textwidth]{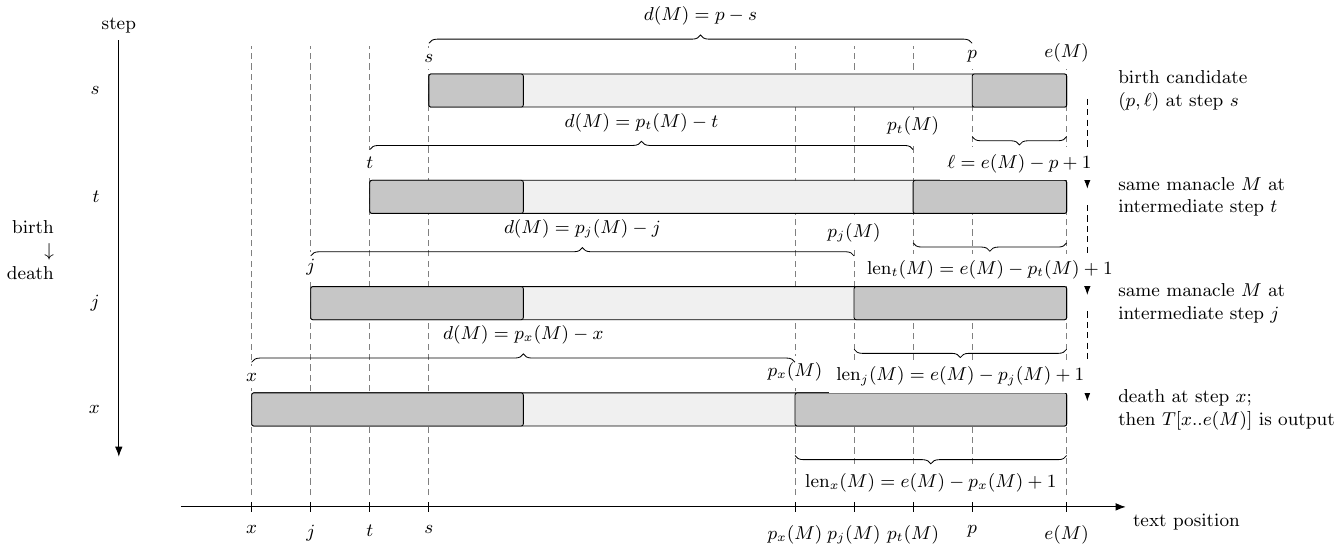}
  \caption{A sketch of a manacle. $x$, $j$, $t$, and $s$ are shown both as step labels and as positions. The invariant gap $d(M)$ is visible as the horizontal distance from the current start position to the matching occurrence of the second border. In each row, the two dark arms have equal length and are separated by a gap larger than an arm.}
  \label{fig:manacle}
\end{figure}

Retrieving \emph{all} candidates $\mathcal C_{j}$ at position $j$ and checking whether each defines an MCS or can be extended to the left would be too time-consuming for an output-sensitive algorithm.
Instead, we group candidates connected by successive left extensions into a manacle, representing them compactly.
Formally, we make a definition grouping maximal sequences of candidates connected by repeated left extension into what we call \emph{manacles}.
See \cref{fig:manacle} for an illustration.

\begin{definition}
\label{def:manacle}
A maximal sequence $M=\{(p,\ell)\in\mathcal C_s,\;(p-1,\ell+1)\in\mathcal C_{s-1},\;\dots,\;(p-(s-x),\ell+(s-x))\in\mathcal C_x\}$ is called a \emph{manacle}. 
We call $d(M):=p-s$ its \emph{gap} and $e(M):=p+\ell-1$ its \emph{right end}.
\end{definition}

At step $j$, a manacle represents the candidate with
label $p_j(M)=j+d(M)$ and length $\mathrm{len}_j(M)=e(M)-p_j(M)+1$. 
Step $s$ is called the \emph{birth} and step $x$ the \emph{death} of $M$.
We say that the candidates $(p,\ell), (p-1, \ell+1), \dots, (p-(s-x),\ell+(s-x))$ are \emph{locked} in $M$.
In particular, the manacle locking a candidate is uniquely determined by the candidate's label and length:

\begin{lemma}
\label{lem:death}
Let $(q,r)\in\mathcal C_s$ and $x=s-\LCS(s-1,q-1)$. 
Then the manacle locking $(q,r)$ has death $x \ge 1$, 
and the MCS obtained at its death is $T[x..q+r-1]$.
\end{lemma}

\begin{proof}
Let $M$ be the manacle locking $(q,r)$. 
By \cref{lemma:manacle}, 
$M$ continues from step $t$ to step $t-1$ if and only if
$T[t-1]=T[p_t-1]$, where $p_t=q-(s-t)$. Walking to the left, the compared characters are
$T[s-1],T[s-2],\dots$ against $T[q-1],T[q-2],\dots$, so the number of successful steps is exactly
$\LCS(s-1,q-1)$, after which either a mismatch occurs or position $0$ is reached, where the sentinel
$\$_1$ differs from every character of $\Sigma$. 
Since $\LCS(s-1,q-1)\le s-1$ we get $x\ge1$. 
The right end $e(M)=q+r-1$ is invariant along $M$. At step $x$, the left extension fails (or the left sentinel is reached), so \cref{lemma:manacle} identifies $T[x..q+r-1]$ as an MCS.
\end{proof}

\section{Algorithm}

In a preprocessing phase we build:
\begin{itemize}
\item a data structure supporting constant-time $\LCE$ and $\LCS$ queries: 
      for instance from the suffix array of $T$ and of the
      reverse of $T$, each with its LCP array and a range-minimum structure
\item the array $\mathrm{next}[1..n]$ with $\mathrm{next}[i]=\min\{q>i:T[q]=T[i]\}$, obtained by a
      single right-to-left scan. It answers $\nxt_i(1)$ in constant time
\item a weighted level ancestor structure on the suffix tree of $T$ (equivalently, a search for the
      suffix-array interval of a prefix of a given suffix), which returns the suffix-array interval of
      $T[i..i+h-1]$ from $i$ and $h$
\item a range successor structure on the suffix array, which returns the smallest text position
      greater than a given threshold inside a given suffix-array interval.
\end{itemize}
The last two together evaluate $\nxt_i(h)$ for $h\ge2$.
By $t_{\text{nxt}}$ we denote the time cost of one such evaluation. 
In addition, the algorithm maintains a doubly linked list $\mathcal L$ of manacles ordered by length, 
and buckets $\mathcal D[1..n]$ of manacles indexed by death step, where $\mathcal D[i]$ stores in an (unordered) bucket the manacles that die at step $i$.

\begin{figure}[htpb]
  \centering
  \includegraphics[width=0.8\textwidth]{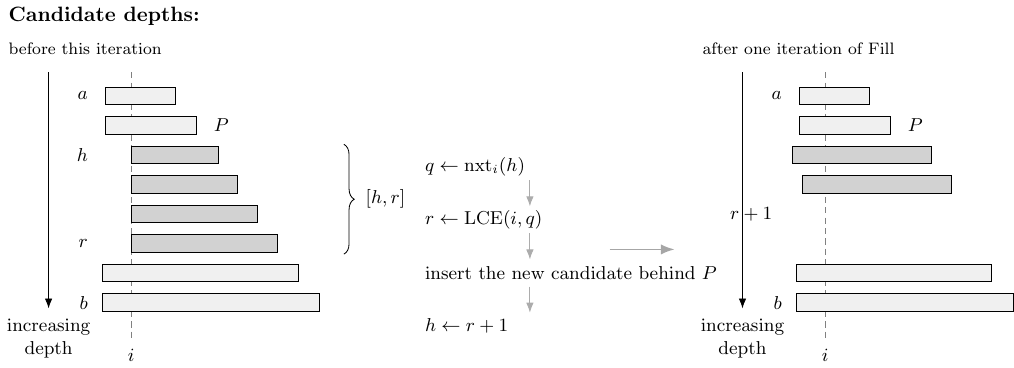}
  \caption{Setting for calling $\textsf{fill}(i,a,b,P)$. 
  Light-gray candidates already start before $i$. 
After the insertion, 
the new candidate for $[h,r]$ also starts before $i$, 
and the unresolved part is shown as a gap.
  }
  \label{fig:fill}
\end{figure}

\cref{alg:manacles} scans the text from right to left, from $j=n$ down to $j=1$.  
The explicit test at Line~\ref{line:char} at the start of each iteration reports the MCSs of length one.
All longer MCSs are handled through manacles.  At the beginning of the iteration for $j$, the list $\mathcal L$ represents the candidates of $\mathcal C_j$ in increasing order of length. 
A manacle in $\mathcal D[j]$ dies at this step and yields an MCS starting at $j$.
Every other manacle survives and, by \cref{lemma:manacle}, automatically represents the corresponding candidate of $\mathcal C_{j-1}$: although its label decreases by one and its length increases by one, the stored values $d(M)$ and $e(M)$ remain unchanged.
Thus, the transition from $\mathcal C_j$ to $\mathcal C_{j-1}$ only has to construct the \emph{new} candidates that are not represented by surviving manacles.

This is the role of \Fill{}, a procedure for finding new manacles.
Removing a maximal consecutive block $f,\ldots,g$ of dying manacles leaves
an interval $[a,b]$ of candidate lengths that may have to be filled at step $j-1$.  
Starting from a manacle $M$ in the doubly linked list $\mathcal L$, we scan left and right to find the maximal consecutive block of dying manacles containing $M$,
denoted $f,\ldots,g$.
Let $P$ be the surviving predecessor of the block, if it exists, so newly created manacles must be inserted directly
behind $P$.
If $P$ exists then its new length is $\mathrm{len}_j(P)+1$, and therefore the first potentially missing length is $a=\mathrm{len}_j(P)+2$.
If the block reaches the head of the list we instead take $a=1$. The upper endpoint is $b=\mathrm{len}_j(g)+1$. The additional call \Fill{$j-1,1,1,\bot$} handles depth~$1$ when the first manacle survives and no deleted block covers it.

Inside \Fill{$i,a,b,P$}, the variable $h$ is the first depth after the candidate represented by $P$
(or $h=1$ if $P=\bot$).  The query $q=\nxt_i(h)$ finds the leftmost succeeding occurrence of length~$h$.
If $q=\infty$, \Fill{} returns, since no longer prefix has a succeeding occurrence.
Otherwise, $r=\LCE(i,q) \ge h$ is the length of the candidate whose depth range starts at $h$. Now:

\begin{itemize}
\item
  If $r \le b$, this candidate is missing and is created.
  By \cref{lem:staircase}, its depth range is $[h,r]$,  so the scan can jump straight to $h=r+1$.  
  The new manacle is inserted after $P$ and put into the bucket $i-\LCS(i-1,q-1)$ where it will die.
  We continue with $P$ being the newly created manacle.
\item
  If $r>b$, then by construction there is already a surviving manacle for this candidate and we can simply return.
\end{itemize}

\begin{algorithm}[h]
\caption{Enumeration of all MCSs of $T[1..n]$.
$\mathrm{len}_j(g)$ gives the length of the candidate of the manacle $g$ at step $j$; 
$\mathrm{pred}(f)$ gives the entry preceding $f$ in $\mathcal L$;
$\mathrm{succ}(f)$ gives the entry following $f$ in $\mathcal L$.
}\label{alg:manacles}
\DontPrintSemicolon
\SetKwProg{Pr}{Procedure}{:}{}
\KwData{$\mathcal L$: doubly linked list of manacles ordered by length; $\mathcal D[1..n]$: buckets}
$\mathcal L\gets\langle\,\rangle$;\quad $\mathcal D[j]\gets\langle\,\rangle$ for all $j$\;
\For{$j=n,n-1,\ldots,1$}{
  \tcp{invariant: $\mathcal L$ represents $\mathcal C_j$, ordered by length (\cref{thm:inv})}
  \lIf{$T[j-1]\neq T[j]$ and $T[j]\neq T[j+1]$}{\textbf{output} $(j,j)$ (easy -- singleton MCS)}\label{line:char}
  \lForEach{$M\in \mathcal D[j]$}{\textbf{output} $(j,e(M))$;\quad mark $M$ as \emph{dying}}
  \lIf{$j=1$}{\Return}
  $\mathit{topDone}\gets\textbf{false}$\;
  \ForEach{$M\in \mathcal D[j]$ that is still in $\mathcal L$}{
    $f\gets M$;\quad \lWhile{$\mathrm{pred}(f)$ is dying}{$f\gets\mathrm{pred}(f)$}
    $g\gets M$;\quad \lWhile{$\mathrm{succ}(g)$ is dying}{$g\gets\mathrm{succ}(g)$}
    $P\gets\mathrm{pred}(f)$;\quad $b\gets\mathrm{len}_j(g)+1$\;
    \lIf{$P=\bot$}{$a\gets1$;\quad $\mathit{topDone}\gets\textbf{true}$}
    \lElse{$a\gets\mathrm{len}_j(P)+2$}
    remove $f,\dots,g$ from $\mathcal L$\;
    \Fill{$j-1,a,b,P$}\;
  }
  \lIf{not $\mathit{topDone}$}{\Fill{$j-1,1,1,\bot$}}
}
\BlankLine
\Pr{\Fill{$i,a,b,P$}}{
  $h\gets a$\;
  \While{$h\le b$}{
    $q\gets\nxt_i(h)$\;
    \lIf{$q=\infty$}{\Return}
    $r\gets\LCE(i,q)$\;
    \lIf{$r>b$}{\Return}
    create a manacle $M$ with $d(M)\gets q-i$ and $e(M)\gets q+r-1$\;
    insert $M$ into $\mathcal L$ directly behind $P$ (at the head if $P=\bot$);\quad
      $P\gets M$\;
    insert $M$ into $\mathcal D[\,i-\LCS(i-1,q-1)\,]$\;
    $h\gets r+1$\;
  }
}
\end{algorithm}

We now prove that calling $\Fill(i,a,b,P)$ finds the missing manacles, and inserts them at the right place in $\mathcal L$.
\begin{lemma}
\label{lem:fill}
Consider a call $\Fill(i,a,b,P)$ and assume
\begin{itemize}
\item[(F1)] either $a=1$ and $P=\bot$, or $P$ is a manacle of $\mathcal L$ whose candidate at
      step $i$ has length $a-1$
\item[(F2)] no candidate of $\mathcal C_i$ whose length lies in $[a,b]$ is already represented in
      $\mathcal L$
\item[(F3)] $b\ge \lambda_i$, or the candidate of $\mathcal C_i$ whose depth range contains $b+1$ is
      already represented in $\mathcal L$.
\end{itemize}
Then the call inserts into $\mathcal L$
those candidates of $\mathcal C_i$ whose length lies in $[a,b]$ in increasing order of length directly behind $P$.
\end{lemma}

\begin{proof}
At the start of each loop iteration, either $h=1$ and $P=\bot$, or $P$ represents a candidate of $\mathcal C_i$ of length $h-1$.
Additionally, the candidates of $\mathcal C_i$ with lengths in $[a,h-1]$ have already been inserted by this call. This holds initially by (F1), and after inserting a candidate of length $r$ it is preserved by setting $P$ to that candidate and $h=r+1$.

Let $q=\nxt_i(h)$. If $q=\infty$, then $h>\lambda_i$, so no candidate has its length in $[h,b]$
and returning is correct. Otherwise put $r=\LCE(i,q)$. By the loop invariant and
\cref{lem:staircase}, $(q,r)\in\mathcal C_i$ is the candidate immediately following $P$ (or the first candidate if $P=\bot$), and its depth range is $[h,r]$. If $r\le b$, then $r\in[a,b]$, so by (F2) this candidate is not yet represented. 
The algorithm inserts the manacle of the candidate~$(q,r)$ directly behind $P$, and continues with $h=r+1$.
If $r>b$, then the depth range $[h,r]$ contains $b+1$.
So $b<\lambda_i$, and by (F3) this same candidate is already represented in $\mathcal L$.
We conclude that no candidate with length in $[h,b]$ is missing and returning is correct.
\end{proof}

Note that every iteration of the while-loop of \Fill{} (except possibly the last) inserts one candidate, which means that \Fill{} evaluates $\nxt_i$ at
most $z+1$ times, where $z$ is the number of inserted candidates.

\begin{theorem}
\label{thm:inv}
For every $j=n,n-1,\dots,1$, at the beginning of the iteration for $j$ the list $\mathcal L$ contains
exactly one manacle per candidate of $\mathcal C_j$, ordered by increasing length, and each such manacle
$M$ satisfies $p_j(M)=p$ and $\mathrm{len}_j(M)=\ell$ for its candidate $(p,\ell)$. Moreover, every manacle of $\mathcal L$ is stored in the bucket $\mathcal D[x]$ for its death step $x$.
\end{theorem}

\begin{proof}[Proof via induction on decreasing $j$.] 
\textbf{Base case:}
For $j=n$ we have $\mathcal C_n=\emptyset$, since no $q>n$ exists, and
$\mathcal L$ is empty.

\textbf{Induction hypothesis:}
Assume the statement holds at step $j>1$, and consider the transition performed in iteration $j$ to establish it at step $j-1$. 

\textbf{Inductive step:}
By the induction hypothesis and \cref{lem:death}, the manacles in $\mathcal D[j]$ are exactly the manacles of $\mathcal L$ whose candidate $(p,\ell)$ is left-maximal, i.e., satisfies $T[j-1]\neq T[p-1]$.
We call these \emph{dying} and all others \emph{surviving}. For each such manacle $M$, the algorithm outputs $(j,e(M))$, which is an MCS by \cref{lemma:manacle}.

By \cref{lemma:manacle}, every surviving manacle $M$ represents a candidate of $\mathcal C_{j-1}$ with label $p-1$ and length $\ell+1$, so its stored pair $(d(M),e(M))$ needs no change and the relative order by length is preserved. 
Conversely, let $(p',\ell')\in\mathcal C_{j-1}$. If $\ell'=1$, its length lies in
the interval $[1,1]$.
If $\ell'\ge2$, let $p=\nxt_j(\ell'-1)$. Then by \cref{lem:shift} either $p'=p-1$, which happens exactly when the candidate of $\mathcal C_j$ whose depth range contains $\ell'-1$ survives, or $p'>p-1$.
In the first case $\ell'-1$ must be the length of that surviving candidate, as if $\ell'-1$ were strictly inside its depth range, then $\LCE(j-1,p')=1+\LCE(j,p)>\ell'$, a contradiction. 
So the candidates of $\mathcal C_{j-1}$ split into the shifted surviving ones and those whose length lies in
\[
  H=\bigl(\{1\}\cup\{\ell+1 : \ell \text{ lies in the depth range of a dying candidate}\}\bigr)\cap[1,\lambda_{j-1}].
\]

Because the depth ranges partition $[1,\lambda_j]$ (\cref{lem:staircase}) and $\lambda_{j-1}\le \lambda_j+1$, the union of the intervals
$[a,b]$ of the calls of \Fill{} in iteration $j$, intersected with $[1,\lambda_{j-1}]$, is exactly $H$:
a maximal block $f,\dots,g$ of consecutive dying manacles contributes
$[\mathrm{len}_j(P)+2,\;\mathrm{len}_j(g)+1]$ for its surviving predecessor $P$, or
$[1,\mathrm{len}_j(g)+1]$ if there is none, and if the first manacle of $\mathcal L$ survives then
depth~$1$ is covered by the final call $\Fill(j-1,1,1,\bot)$.

We verify the conditions of \cref{lem:fill}. For a call belonging to a maximal dying block $f,\dots,g$, condition (F1) holds because, if $P$ exists, it survives and its length at step $j-1$ is $\mathrm{len}_j(P)+1=a-1$; if it does not exist, then $a=1$ and $P=\bot$.
Condition (F2) holds because the surviving predecessor has length $a-1$ and the surviving successor (if any) has length strictly larger than $b$, meaning that all surviving candidates lie outside $[a,b]$.
Additionally, the intervals contributed by distinct maximal dying blocks are disjoint, so candidates inserted by earlier calls lie outside $[a,b]$ as well. Thus no candidate with length in $[a,b]$ is represented when the call
is made.

For (F3), suppose first that $N=\mathrm{succ}(g)\neq\bot$. Then $N$ survives. Since
$b=\mathrm{len}_j(g)+1$ is the first depth in the depth range of $N$ at step $j$, we have
$\nxt_j(b)=p_j(N)$. Applying \cref{lem:shift} at depth $b+1$ gives
$\nxt_{j-1}(b+1)=p_j(N)-1$. 
Hence, $b+1$ lies in the depth range of the candidate represented by $N$
at step $j-1$, as required by (F3). If $N=\bot$, then
$b=\lambda_j+1\ge \lambda_{j-1}$ by \cref{lem:shift}, so (F3) again holds.

It remains to check the additional call $\Fill(j-1,1,1,\bot)$. It is made only when no dying block reaches the head of the list. Thus either $\mathcal C_j$ was empty, in which case $\lambda_j=0$ and $\lambda_{j-1}\le \lambda_j+1=1$, or the first manacle survives. In the latter case no represented candidate at step $j-1$ has length~$1$, so (F2) holds.
And if $\lambda_{j-1}>1$, \cref{lem:shift} applied at depth~$2$
to the first surviving candidate shows that depth~$2$ lies in its depth range. Thus (F1)--(F3) hold also for this call.

By \cref{lem:fill}, all calls insert only the candidates of $\mathcal C_{j-1}$ with length in $H$, each once and at their correct position.
Together with the surviving manacles, $\mathcal L$ therefore represents $\mathcal C_{j-1}$, ordered by length.

Finally, note that every newly created manacle is inserted into the bucket of its death step, and that the dying manacles are removed from $\mathcal L$.
\end{proof}

\begin{theorem}
\label{thm:complete}
\cref{alg:manacles} outputs exactly the MCSs of $T$.
\end{theorem}

\begin{proof}
MCSs of length one are reported in Line~\ref{line:char} in \cref{alg:manacles}.
For each $i$, \cref{thm:inv} gives exactly one manacle for every $(p,\ell)\in\mathcal C_i$. By \cref{lem:death}, this manacle dies at step~$i$ if and only if $T[i-1]\neq T[p-1]$.
In this case the algorithm reports $(i,p+\ell-1)$.
By \cref{cor:char}, these are the MCSs of length at least~$2$ starting at~$i$. Finally, by \cref{lem:shape}, each such MCS determines a unique $\ell$, hence a unique candidate $(p,\ell)$, and therefore is reported exactly once.
\end{proof}

\begin{theorem}
\label{thm:time}
After preprocessing, \cref{alg:manacles} runs in time $O(n+k)$ plus $O(k)$ evaluations of $\nxt_i(h)$ with $h\ge2$, which results in time $O(n+k t_{\text{nxt}})$.
It uses $O(n)$ words of working space (in addition to the range successor data structure).
\end{theorem}

\begin{proof}
By \cref{thm:complete} every manacle is output as exactly one MCS at its death, so at most $k$ manacles are ever created. Each manacle is created, inserted, output and removed once, at constant cost each, and the scan of a maximal block of dying manacles is charged to its members.
Apart from these charges, each of the $n$ iterations costs $O(1)$ time.

To bound the successor queries, let $c$ be the number of created manacles and $\beta$ the number of \Fill{} calls associated with dying blocks. Every such block is nonempty, and each manacle dies only once, so $\beta\le c\le k$.
By \cref{lem:fill}, these calls jointly evaluate $\nxt$ at most $c+\beta\le 2k$ times.
The only remaining \Fill{} calls have parameters $(j-1,1,1,\bot)$.
Each of them evaluates $\nxt_{j-1}(1)$ at most once, answered in $O(1)$ time by $\mathrm{next}$. 
So at most $2k$ evaluations have $h\ge2$ and cost $t_{\text{nxt}}$ each, while all $\LCE$ and $\LCS$ queries take constant time.

For the space bound, all preprocessing structures other than the range-successor structure occupy $O(n)$ words, and the only other storage is the bucket array $\mathcal D[1..n]$, again $O(n)$ words, together with the manacles.
The record of a manacle $M$ is created when $M$ is born and is released as soon as $M$ dies, since a dead manacle is never inspected again.
By \cref{thm:inv}, at the beginning of iteration $j$ the manacles in use are exactly those of the $|\mathcal C_j|$ manacles representing $\mathcal C_j$, and during iteration $j$ the records of the dying manacles are still needed for the scan of $\mathcal D[j]$ while the manacles of $\mathcal C_{j-1}$ are already
being created, so at most $|\mathcal C_j|+|\mathcal C_{j-1}|$ records are in use at any instant.
This is at most $n$ by \cref{lem:width}.

\end{proof}

\begin{corollary}
\label{cor:concrete}
For every constant $\varepsilon>0$, all $k$ MCS occurrences of a string $T[1..n]$
over a polynomially bounded integer alphabet can be enumerated deterministically
with either of the following bounds in the word-RAM model:
\begin{enumerate}
\item $O(n\sqrt{\log n}+k\log^\varepsilon n)$ total time and $O(n)$ words of space, or
\item $O(n\sqrt{\log n}+k\log\log n)$ total time and
      $O(n\log\log n)$ words of space.
\end{enumerate}
Both time bounds include preprocessing, and each MCS is reported by its pair of endpoints.
\end{corollary}

\begin{proof}
The suffix-array interval of a text fragment $T[i..i+h-1]$ can be located in $O(1)$ time using the suffix-tree weighted-ancestor structure of Belazzougui et al.~\cite{belazzougui2021weighted}, which takes $O(n)$ words and $O(n)$ construction time. 
Thus, $\nxt_i(h)$ reduces to a range successor
query on the suffix array, with threshold $i$, in addition to $O(1)$ work.

By Proposition~9.7(a,b) of Kociumaka et al.~\cite{kociumaka2024internal},  based on the range-successor structures of Belazzougui and Puglisi~\cite{belazzougui2016range} and Gao et al.~\cite{gao2020range}, range successor queries can be supported with either $O(n)$ words and $O(\log^\varepsilon n)$ query time, or $O(n\log\log n)$ words and $O(\log\log n)$ query time.
Both structures have $O(n\sqrt{\log n})$ construction time. 
The other preprocessing structures are constructed in $O(n)$ time and space. 
Substituting these bounds into \cref{thm:time} proves the two claims.
\end{proof}

\section{Future Work}
\label{sec:future-online}

A natural direction is faster \emph{reverse-online} enumeration, 
with characters arriving from right to left.  
The MCSs beginning at $i$ are reported once $T[i-1]$ becomes known, 
allowing a one-character delay to test left maximality.  
For a constant alphabet and known final length $n$, we aim for $O(n)$ words and amortized total time
\(
  O\!\left((n+k)\frac{\log n}{\log\log n}\right).
\)

Our idea is to represent the candidate staircase visualized in \cref{fig:staircase} within the suffix tree, similar to \cite{shibata2026online}.
In the suffix tree $\mathcal T_i$ of $T[i..n]\$_2$, 
label each vertex $v$ by $\nu_i(v)$, 
the smallest later suffix position $q>i$ below $v$, or $\infty$ if none exists.  
Along the path to leaf $i$, the label at string depth $h$ is $\nxt_i(h)$.  
Thus a label-changing edge $v\to u$ with parent string depth $h>0$ identifies the candidate $(q,h)$, 
where $q=\nu_i(v)$; 
see \cref{lem:staircase}.
By \cref{cor:char}, this edge witnesses the MCS $(i,q+h-1)$ if and only if $T[i-1]\ne T[q-1]$, with the left sentinel used at $i=1$.
Representing the information $T[q-1]$ as the color of this edge,
reporting amounts to color-filtered enumeration of label-changing edges on a root-to-leaf path.

When $T[i-1]$ arrives, the newly eligible later position $i$ becomes the minimum label of every old vertex on the root-to-leaf-$i$ path; 
other old labels remain unchanged.  
After this \emph{root-path assignment}, we insert the new suffix.  

A possible implementation uses a recursive \emph{macro/micro-tree decomposition} of the suffix tree. 
Microtrees of $M=\Theta((\log n)^{1/4})$ items would store packed path summaries,
including endpoint labels and colors of label changes.  
Contracting microtrees recursively gives $H=O(\log n/\log\log n)$ levels.  
With constant-time local packed operations, 
a path update or a report visits $O(H)$ fragments, 
apart from fragments charged to its outputs.
An oversized microtree could be rebuilt after splitting: 
charging $O(MH)$ rebuilding work to $\Omega(M)$ insertions would suffice for an $O(H)$ amortized update cost.
The unresolved part is the dynamic update of the recursively defined macro/micro-tree decomposition.

\end{document}